\documentstyle[pre,aps,multicol,eqsecnum,epsf,rotate]{revtex}
\begin{document}
\draft
\tighten

\newcommand{\r}[1]{(\ref{#1})}                       
\newcommand{\p}[1]{\left( #1\right)}                 
\newcommand{\bm}[1]{\mbox{\boldmath $#1$}}           


\title
{\Large Dynamic Light Scattering from Semidilute Actin Solutions: \\
A Study of Hydrodynamic Screening, Filament Bending Stiffness \\
and the Effect of Tropomyosin/Troponin-Binding }
\author{R. G\"otter, K. Kroy, E. Frey, M. B\"armann, and E. Sackmann}
\address{Department of Physics
 E22 (Biophysics Group), \\
Department of Physics T34 (Theoretical Physics), \\
Technische Universit\"at M\"unchen, \\
James-Franck-Strasse, D-85747 Garching, Germany.}
\date{\today}
\maketitle

\begin{abstract}
  Quasi-elastic light scattering (QELS) is applied to investigate the effect of
  the tropomyosin/troponin complex (Tm/Tn) on the stiffness of actin filaments.
  The importance of hydrodynamic screening in semidilute
  solutions is demonstrated.  A new concentration dependent expression for the
  dynamic structure factor $g(\bm k,t)$ of semiflexible polymers in semidilute
  solutions is used to analyze the experimental QELS data. A concentration
  independent value for the bending modulus $\kappa$ is thus obtained. It
  increases by 50\% as a consequence of Tm/Tn binding in a 7:1:1 molar ratio of
  actin/Tm/Tn.  In addition a new expression for the
  initial slope of the dynamic structure factor of a semiflexible polymer is
  used to determine the effective hydrodynamic diameter of the actin filament.
  Our results confirm the general relevance of the concept of (intrinsic)
  semiflexibility to polymer dynamics.
\end{abstract}

\begin{multicols}{2}

\section{Introduction}
\begin{figure}
  \narrowtext \epsfxsize=0.9\columnwidth \epsfbox{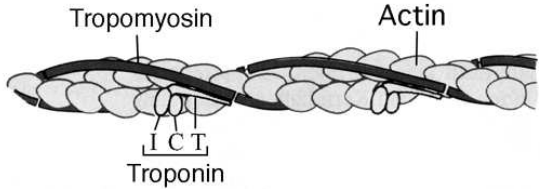}
  \vspace{0.2truecm}
  \caption{Model of Ca$^{++}$ mediated Tm/Tn binding to actin filaments.
    Tm/Tn is known to bind along the groove of the actin filaments.}\label{bin}
\end{figure}

The main motivation of the present work was to study the effect of
tropomyosin/troponin (Tm/Tn) binding on the bending stiffness of actin
filaments. The semiflexible filaments with a length of 400\,\AA\ bind along the
groove between the two twisted strands of the actin filament
(Figure~\ref{bin}).  The binding is influenced by the presence of Ca$^{++}$.
This effect is of great biological significance, since it plays a central role
in the regulation of the myosin-actin coupling in muscles.

\begin{figure}[tb]
  \narrowtext \hspace{0.005\columnwidth} \epsfxsize=0.9\columnwidth
  \epsfbox{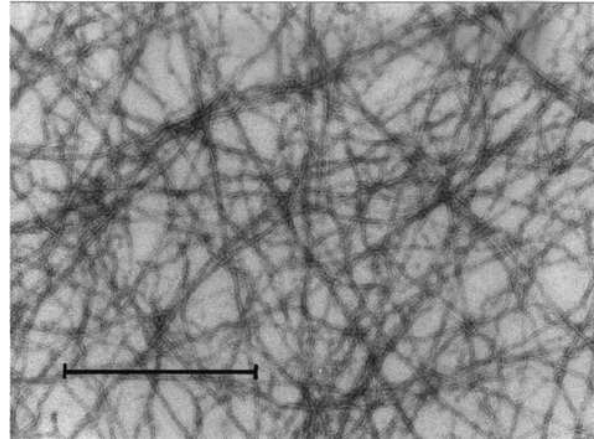} \vspace{0.2truecm}
\caption{Electron micrograph of a \protect $0.4 \frac{\rm mg}{\rm ml}$
  actin solution polymerized in vitro. The bar indicates the length of 1
  $\mu$m. We probe the samples with scattering wavelengths \protect
  $\lambda=2\pi/k$ that are somewhat shorter than the mesh size \protect $\xi$
  of the network, which, in turn, is shorter than the persistence length
  \protect $L_p$. The global motion of all polymers is strongly hindered by the
  entanglement.}\label{mic}
\end{figure}

Actin, when polymerized in vitro, forms semiflexible macromolecules of contour
lengths $L$ up to some 30 $\mu$m and with a persistence length $L_p$ of at
least some $\mu$m. The system we are interested in is a semidilute solution of
these macromolecules, i.e.\ an entangled network, where the single actin
filaments interact strongly but the free volume is still much larger than the
excluded volume.  As a consequence of the great extension and large persistence
length of the molecules, as compared to their lateral diameter $a$ (some nm),
the semidilute regime is unusually large.  We probe the samples with scattering
wavelengths $\lambda=2\pi/k$ that are somewhat shorter than the mesh size
$\xi$ of
the network, which, in turn, is shorter than the persistence length $L_p$;
i.e.\ the condition
\begin{equation}\label{skaltrenn}
a\ll\lambda < \xi \ll L_p,\,  L
\end{equation}
is fulfilled for all of our samples.  An electron micrograph of a typical
system under study is reproduced in Figure~\ref{mic}.

The bending stiffness of a filament-like macromolecule may be inferred from an
analysis of its conformational dynamics.  Therefore various experimental
techniques have been applied to investigate the dynamical properties of actin
\cite{jan94}.  Solutions and gels were probed by high sensitivity rheology
using torsional \cite{mul91} and magnetic bead rheometers \cite{zie94}. K\"as
et al.\ \cite{kas93} and others \cite{ott93} analyzed single labeled filaments
by microfluorescence microscopy combined with dynamic image processing.  There
have also been several attempts to establish QELS as a quantitative method to
probe the static and dynamical properties of semiflexible macromolecules (for
critical discussions of the literature see e.g.\ Refs.\
\cite{son91,ara91,har95}).  But in contrast to the dynamic properties of
flexible polymers the dynamics of semiflexible polymers is not yet very well
understood. This is mainly a consequence of the difficulties caused by the
rigid constraint of a (virtually) unextensible contour. Models that try to
represent the unextensible contour honestly \cite{ara85,ara91} have to deal
with considerable technical difficulties. On the other hand, models that relax
the constraint like the various modified versions of the so called
Harris-Hearst-Beals-model \cite{har66} (for a short summary see \cite{har94})
allow for artificial stretching modes and predict a gaussian probability
distribution for the spacial distances of the contour elements. As the most
promising model with relaxed constraint we regard the model described in
Refs.\
\cite{lag91,har94,har95}.  We will neither relax the constraint, nor do we need
all the machinery developed by Arag\'on and Pecora in Refs.\
\cite{ara85,ara91}.  (However, one of our results, Eq.\ \r{res}, may be
obtained by their method \cite{krotp}.)  In the special case, we are interested
in, we profit from some simplifications brought about by the separation of
length scales, Eq.\ \r{skaltrenn}, which in turn gives rise to a hierarchy of
time scales: QELS measures the temporal decay of configurational correlations
of the filaments.  For long semiflexible actin filaments in entangled networks
the internal configurational dynamics dominates over the center of mass and
rotational motion of the molecules in time intervals typically probed by QELS.
In addition, we can restrict ourselves to the weakly bending rod limit, where
one can approximately take the undulations to be transverse to the mean contour
\cite{mae84,son91}, which is virtually fixed for the relevant time scales.  The
decay rate of the dynamic structure factor becomes thus accessible to a simple
physical interpretation in terms of the bending modes only, i.e.\ the local
bending modes may be studied ``in isolation". This is a great practical
advantage compared to the general case.

But the time decay of the structure factor also depends on the strength of
hydrodynamic correlations.  To obtain quantitative results it is therefore
necessary to take into account the screening of the hydro\-dynamic
self-interaction of the filaments in semidilute solutions (or their finite
length in the dilute case). As explained in Section~\ref{str} this is most
simply achieved by means of a concentration dependent renormalized friction
coefficient, $\zeta_{\perp}$.

In contrast to the expression for the dynamic structure factor for intermediate
times, which takes a simple form only after some approximations, its initial
slope, resulting from the quasi-free Brownian fluctuations about the
equilibrium configuration may be calculated exactly if Eq.\ \r{skaltrenn}
holds.  The result is compared with previous experimental data in
Section~\ref{ins}. The good agreement of the theoretical prediction with the
available data confirms our hypothesis that the intrinsic rodlike structure of
polymers is a significant detail, which influences the dynamic properties.
Moreover, the microscopic hydrodynamic diameter $a$ of the filament, which
enters the dynamic structure factor as a parameter, also appears in the
expression for the initial decay rate and may thus be determined by a
measurement of the initial slope. Section~\ref{mat} lists materials and
methods, and the new experimental results are presented in Section~\ref{exp}.
Finally we discuss possible refinements to our approach in Section~\ref{dis}.

\section{Theoretical Background of QELS from Semiflexible Polymers}\label{the}

\subsection{Stretched Exponential Decay of the Dynamic Structure
  Factor}\label{str}

The dynamic structure factor of a chain of length $L=N\Delta s$ with $N$
segments located at $\bm r_n$ ($n=1,\dots ,N$) is defined by
\begin{equation}\label{def} g(\bm{k},t)=\frac1N \sum_{n,m}\langle
\exp[i\bm{k}(\bm{r}_n(t)-\bm{r}_m(0))] \rangle \, .
\end{equation}
The brackets $\langle \dots \rangle$ denote the ensemble average over all chain
conformations, and \bm k is the scattering wave vector.  On the typical length
scales probed by QELS ($0.1\sim 1$ $\mu$m) the conformational dynamics of actin
filaments are dominated by the bending undulations. Consequently, we take the
bending energy, given by the contour integral $\int_Lds$ over the local
curvature multiplied by the bending modulus $\kappa$,
\begin{equation}\label{ene} E[\bm{r}_s]=\frac{\kappa}2 \int_L
  ds\Bigl(\frac{\partial^2 \bm{r}}{\partial s^2}\Bigr)^2 \, ,
\end{equation} to
be the only relevant energetic term in a canonical description.  This implies
that the static mean square end-to-end-distance is given by the well known
Kratky-Porod formula \cite{kra49} with the persistence length $L_p
=\kappa/k_BT$. The time decay of the structure factor $g(\bm{k},t)/g(\bm{k},0)$
is not easily calculated in the most general case. However, the particular case
under study allows for simplifications, which enable us to adapt some of the
ideas common in the theory of flexible polymers \cite{gen67,doi86}.  We start
from the Langevin equation for a single polymer
 \begin{equation}\label{lan}
  \frac{\partial}{\partial t} \bm{r}_s(t)= \int_L ds'\bm H_{\perp}(\bm r_s,\bm
  r_{s'}) \left( -\frac{\delta}{\delta \bm{r}_{s'}} E[\bm{r}_{s'}]
    +\bm{f}_{s'}\right)\, ,
\end{equation}
where $\bm{f}_s$ is the stochastic force (white noise) and $\bm H_{\perp}$ is
an effective mobility matrix, which takes into account the solvent-mediated
self-interaction of the filament. This can be understood in analogy to the
usual Oseen tensor.  However, the effective reduction of the degrees of freedom
due to the rigid constraint of a fixed contour length requires that the local
longitudinal motion be projected out.  In the weakly bending rod limit this is
practically achieved by a suitable choice of coordinates: longitudinal
distances are kept fixed, and the bending undulations are described by
transverse coordinates (cf.\ Eq.~\r{mob} below). This is how one would address
the problem in analogy to the classical Zimm model for flexible polymers, and
how it was attempted previously in Ref.~\cite{far93}. We shall neglect the
center of mass and rotational motion, which are slow compared to the internal
dynamics of the molecule as a consequence of the scale separation Eq.\
\r{skaltrenn}, the entanglement and hydrodynamic screening.

\begin{figure}[tb]
  \narrowtext \hspace{0.005\columnwidth} \epsfxsize=0.9\columnwidth
  \epsfbox{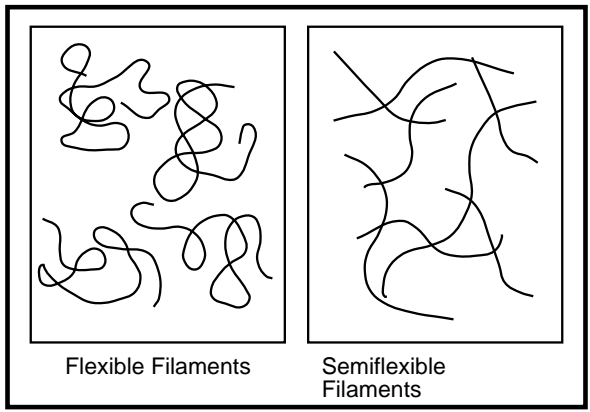} \vspace{0.2truecm}
\caption{Flexible filaments interact mostly with themselves, whereas
  semiflexible filaments interact with each other even at quite low
  concentrations. The perturbation of a single semiflexible filament by its
  surrounding may be modeled by a screening of the hydrodynamic
  self-interaction.}\label{sem}
\end{figure}

For the following it is important to realize that there is a profound
difference between the hydrodynamic interaction of flexible polymers and
semiflexible polymers in semidilute solutions. A flexible polymer is coiled and
is thus far more likely to interact with itself than with surrounding polymers.
On the other hand, semiflexible filaments like actin are much more stretched
and strongly interact with each other even
at quite low concentrations (Figure~\ref{sem}). 
As the scattering vector $\bm k$ in our experiments is large enough in
magnitude to resolve single actin filaments (see Eq.\ \r{skaltrenn}), we are
interested in the dynamics of a single filament and use a mean field
approximation to model the hydrodynamic interaction with the surrounding,
i.e.\ we introduce a screening of the hydrodynamic self-interaction along a
single polymer.  This is achieved by use of the (preaveraged) screened
transverse mobility matrix,
\begin{equation}\label{mob}
\bm H_{\perp}(\bm r) =\frac{e^{- r/\Lambda}}{8\pi\eta r}\left(\bm 1 -
\frac{|\bm{r}\rangle\langle\bm{r}|}{r^2}\right)  ,
\end{equation}
where $\bm r:=\bm r_s-\bm r_{s'}$, $\Lambda$ is the screening length and $\eta$
the solvent viscosity. Beyond this length $\Lambda$, the hydrodynamic
self-interaction of a filament is weak and correlations decay rapidly. The
explicit form of the projector is a consequence of modelling the actin filament
as a straight rod on length scales smaller than $\Lambda$ with respect to the
hydrodynamic interaction.  This should be a good approximation for $\xi \ll
L_p$.

To simplify the calculation we use a kinetic coefficient $\zeta_{\perp}^{-1}$,
\begin{equation}\label{sim}
(\zeta_{\perp}/L)^{-1}=\frac{\log \Lambda/a}{4\pi\eta} \, ,
\end{equation}
in place of the tensor $\bm H_{\perp}$. $\zeta_{\perp}$ is a renormalized
friction, obtained by taking the terms in parentheses in Eq.\ \r{lan} out of
the integral and averaging over all segment positions $s$ for a rigid rod of
diameter $a$. Replacing the mobility matrix by a simple coefficient amounts to
setting the effective friction for all modes equal to the friction of the
dominant mode of wavelength $\Lambda$.  In this approximation interactions
between different modes are neglected and the mobility for the very short
wavelength modes is supposedly slightly overestimated. The latter should not
profit as much by correlations over distances larger than their wavelength, as
is implied by Eq.\ \r{mob}. This approximation will allow for a simple
expression [Eq. \r{res} below] for the structure factor and at the same time
captures the main effect of the hydrodynamic self-interaction of the single
polymers as well as their mutual interaction.  A theoretical basis for the
above ansatz was already worked out by Muthukumar and Edwards \cite{mut83} and
will be explained in more detail in a forthcoming paper \cite{krotp}.

$\Lambda$, $a$ and the persistence length $L_p$ are the three characteristic
length scales in terms of which the dynamics of a semiflexible polymer in
semidilute solution is characterized.  The microscopic cutoff parameter $a$
takes care of the finite thickness of the filament. Its value may be determined
experimentally by the QELS method as described in the next subsection.  We
suppose that the hydrodynamic screening length $\Lambda$ may be identified (up
to a numerical factor) with the mesh size $\xi$ of the actin network,
\begin{equation}\label{mes}
\Lambda \simeq \xi \sim c_a^{-\frac12}.
\end{equation}
(Basically the same relation $\Lambda\sim c^{-1/2}$ can be derived within the
effective medium approach for rods \cite{mut83}.)  It is through Eq.\ \r{mes}
that the actin concentration $c_a$ ultimately enters Eq.\ \r{res} for the
structure factor. The scaling law for the mesh size $\xi$ in Eq.\ \r{mes}
should be valid in the semidilute regime, when $\xi \leq L_p$ and the solution
appears as a random network of almost rodlike segments. This was indeed
confirmed experimentally \cite{sch89}.  The dimensionless quantity $\Lambda/a$,
which could be called the ``hydrodynamic aspect ratio" of the polymer, is
actually the only characteristic parameter of the filament entering expression
\r{res} for the structure factor besides the persistence length $L_p$ (cf.\
Eq.\ \r{dec} below).

Detailed calculations \cite{krotp} show that the relaxation time of the
$p^{th}$ mode is $\tau_p=\frac{\zeta_{\perp}/L}{\kappa}\left(\frac L{\pi p}
\right)^4$, as one would guess from dimensional analysis, and that in the
intermediate time regime\footnote{The second condition is actually
    modified by entanglement. This may indeed be exploited in future high
    precision measurements, as is explained in \protect\cite{krotp}, but is of
    minor importance here.}\label{taux},
\begin{equation}\label{tim}
(kL_p)^{-4/3}\gamma_k^{-1}\ll t\ll \tau_1,
\end{equation}
the structure factor factor is to a good approximation given by a
stretched exponential
\begin{equation}\label{res}
g(\bm{k},t)=g(\bm{k},0)
\exp\left( -\frac{2\Gamma(1/4)}{9\pi} (\gamma_k t)^{3/4} \right).
\end{equation}
The decay rate,
\begin{equation}\label{dec}
\gamma_k=\frac{k_BT}{\zeta_{\perp}/L}k^{8/3}L_p^{-1/3},
\end{equation}
depends on the filament stiffness through the persistence length $L_p$ and on
the actin concentration $c_a$ (or the mesh size) through the renormalized
friction coefficient $\zeta_{\perp}$. It should be compared with the initial
decay rate given below in Eq.\ \r{ini} for the dilute case.  Both are expected
to play an important role in many dynamical problems, i.e.\ they are supposed
to constitute the characteristic time scales of motion.

Some remarks have to be made on the derivation and the domain of validity of
Eq.~\r{res}. First of all it is important to realize that the local motion of
the monomers is neither strictly isotropic (like in the random coil limit) nor
strictly transverse to the mean contour (as was assumed in Eq.\ \r{mob}). The
averaging over the different orientations can be performed in both limits
\cite{krotp}.  Upon neglecting the anisotropy of the local segment motion, one
obtains the simple stretched exponential from of the dynamic structure factor,
Eq. \r{res}.  In the opposite limit -- i.e.\ for local segment motion strictly
transverse with respect to the mean contour -- averaging over the different
orientations is performed after calculating the anisotropic dynamic structure
factor for an individual macromolecule, whose mean orientation and center of
mass position is assumed to be fixed in space and time.  In our case this is
realized by the time scale separation mentioned above and by the fact that the
macromolecules are imbedded in a network. It is found that the resulting
dynamic structure factor is very similar to the stretched exponential form
obtained by neglecting the anisotropy of the segment motion in the averaging
procedure.  We propose to use the stretched exponential form also in the
intermediate regime, where the segment motion is neither strictly transverse
nor isotropic. This allows for a simple fit to the experimental data and has
the advantage that the relevant physical mechanisms are not obscured by
complicated numerical analysis.  But, as a consequence of the approximations
made, the numerical value of the prefactor in the exponential of Eq.~\r{res}
should be taken with some precaution.  Unfortunately, the value obtained for
the persistence length by Eq.~\r{res} is very sensitive to this prefactor as
well as to all the experimental parameters entering the exponent because the
persistence length enters the expression for the decay rate, Eq.\ \r{dec}, as
$L_p^{1/3}$.  In addition the actual prefactor in the scaling law, Eq.~\r{mes},
for $\Lambda$ is not known. Hence the method described here is not capable of
producing very accurate absolute values for the bending modulus so far. On the
other hand, relative changes in the stiffness, which may be caused by diverse
chemical or physical mechanisms are readily detected.

\subsection{Initial Slope of the Dynamic Structure Factor}\label{ins}
\begin{figure}[tb]
  \narrowtext \hspace{0.005\columnwidth} \epsfxsize=0.9\columnwidth
  \epsfbox{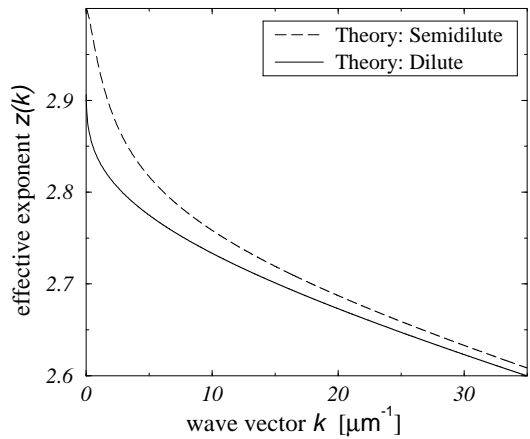}
\caption{Effective exponent $z(k)$ of the ``quasi-scaling"
  law for the initial decay rate of the dynamic structure factor of actin.  $z$
  was predicted to be a universal number, $z=3$, for all flexible polymers by
  classical scale invariant models. This was never observed experimentally.  We
  argue that the (intrinsic) semiflexibility of all real polymers is
  responsible for the discrepancy. The dashed line was computed for a
  semidilute solution of actin filaments ($c_a=0.16\, \frac{\rm mg}{\rm ml}$).
  The solid line corresponds to the dilute case, Eq.\ \protect\r{qsc}. In the
  typical $k$ intervall probed by QELS ($5\sim 30$ $\mu\rm m^{-1}$)
  $z(k)\approx 2.7$ for actin.}\label{zvk}
\end{figure}

Although the dynamic structure factor $g(\bm k,t)$ is a complicated object, and
several physical assumptions and approximations enter its explicit calculation,
its initial slope may be evaluated exactly in the case of Eq.\ \r{skaltrenn}.
The general scheme of computation may be found in Ref.\ \cite{doi86} (see also
Ref.\ \cite{sch84} for a summary of predictions derived from various
semiflexible models).  In the semiflexible case caution has to be paid to the
rigid constraint of constant contour length.  It causes an effective reduction
of the degrees of freedom of the local quasi-free Brownian fluctuations about
the equilibrium configuration, which determine the initial slope for large
scattering vectors and moderate chain stiffness. (For very stiff molecules the
time regime for these quasi-free fluctuations diminishes \cite{son91}, but
actin is far from this limit as can be inferred from a comparison of the
characteristic time scales $\tau_p$, $\gamma_k$ and $\gamma_k^{(0)}$.)  For the
detailed calculation of the initial decay rate,
\begin{equation}\label{gde}
\gamma_k^{(0)} :=-\left. \frac{d}{dt}\log g(\bm k,t)\right|_{t=0}\, ,
\end{equation}
for the case of a semiflexible polymer in semidilute solution we refer the
reader to Ref.\ \cite{krotp}.  Here we give the asymptotic result for a dilute
solution and for scattering wave vectors $\bm{k}$ of modulus $k$ larger than
the inverse persistence length $L_p^{-1}$ but much smaller than the inverse of
the microscopic cutoff length $a$ introduced in Eq. \r{sim}:
\begin{equation}\label{ini}
\gamma_k^{(0)}= \frac{k_BT}{6\pi^2\eta}k^3\left(\frac56 -\log ka\right) \, .
\end{equation}
The deviation from ideal scaling is rather weak, hence it is useful to express
Eq.\ \r{ini} as a ``quasi-scaling law" $\gamma_k^{(0)}\sim k^{z(k)}$ with an
effective dynamic exponent (Figure~\ref{zvk}) given by
\begin{equation}\label{qsc}
z(k)=3\frac{6 \log ka -3}{6 \log ka -5} \, .
\end{equation}
It is quite striking that the microscopic cutoff $a$ appears here.
Hydrodynamic screening does not substantially alter the result of Eq.\ \r{ini}
in the large wave vector regime, but flattens the increase of
$\gamma_k^{(0)}/k^3$ for small wave vectors.  However, the theory is not valid
for small scattering vectors, because scattering vectors smaller than the
inverse mesh size $\xi^{-1}$ do not resolve single filaments but average over
several molecules.  Moreover, as a concequence of local fluctuations in the
mesh size there is scattering in the experimental data in the crossover region
$k\approx \xi^{-1}$.

\begin{figure}[tb]
  \narrowtext \hspace{0.005\columnwidth} \epsfxsize=0.9\columnwidth
  \epsfbox{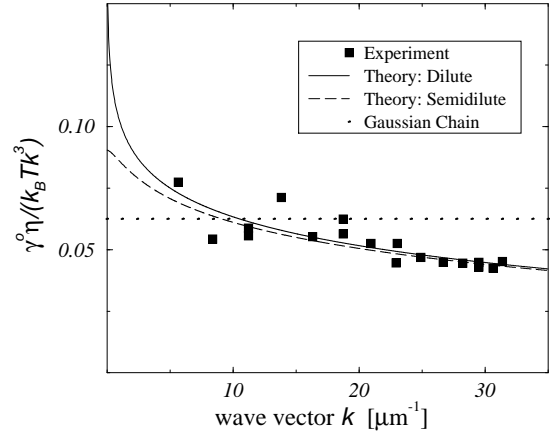}
\caption{Correction to the classical prediction $\gamma_k^{(0)} \sim k^3$ for
  the initial decay rate of the dynamic structure factor. The theoretical
  predictions for dilute solutions (solid line) and semidilute solutions
  (dashed line) are compared with experimental data of Schmidt
  \protect\cite{schdr,sch89}. Also included is the prediction for gaussian
  chains from Ref.\ \protect\cite{doi86}.  The experimental data and the dashed
  line both correspond to the same actin concentration $c_a=0.16\, \frac{\rm
    mg}{\rm ml}$.  The positions of the theoretical curves depend on the
  effective hydrodynamic diameter $a$ of the filament [see Eq.\
  \protect\ref{ini}]. It was used as fit parameter and found to be $a=5.4$
  nm.}\label{cor}
\end{figure}

Figure~\ref{cor} shows a comparison of the computed initial decay rate
$\gamma_{k}^{(0)}$ for a solution of $0.16\, \frac{\rm mg}{\rm ml}$ actin
(dashed line) with previously measured data \cite{schdr,sch89}. A simple least
square fit of Eq.\ \r{ini} to the data gives $a = 5.4$ nm.  $a$ represents an
effective hydrodynamic diameter of the filament to be compared with two times
the cross sectional radius of gyration $r_{\perp}^g$.  The remarkable agreement
of $a$ with the value of $2 \, r_{\perp}^g=5.16 \pm 0.3$ nm determined by
different methods \cite{bre91,ege84} strongly supports the above ideas.

\begin{figure}[tb]
  \narrowtext \hspace{0.005\columnwidth} \epsfxsize=0.9\columnwidth
  \epsfbox{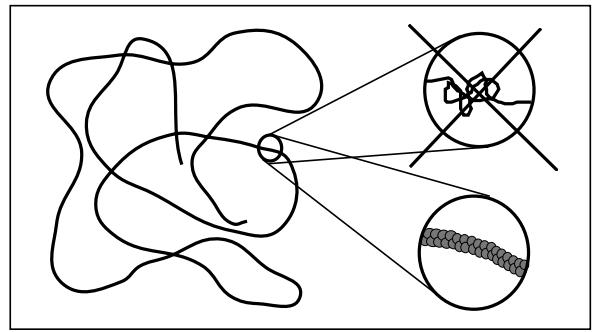} \vspace{0.2truecm}
\caption{A real polymer is not a fractal (i.e.\ not a gaussian chain,
  as often assumed for computational convenience) but rodlike on small scales.
  This affects the dynamical properties of polymers.}\label{rod}
\end{figure}

Finally we would like to note that in classical neutron scattering experiments
with more or less flexible synthetic polymers the ratio of wavelength to
persistence length $\lambda/L_p$ is not very different from what is
encountered in light scattering from large biomolecules.  So one expects
similar results in both cases.  It is well known \cite{gensc,doi86} that there
is sometimes poor agreement between the experiments with synthetic polymers and
the classical theoretical predictions for the initial slope of $g(\bm{k},t)$
derived from scale invariant models.  The excellent agreement achieved now with
actin networks provides strong evidence that the usual scale invariant polymer
models are not capable of describing quantitatively the dynamic properties of
real polymers, which, after all, are semiflexible at heart (Figure~\ref{rod}).

\section{Materials and Methods}\label{mat}

Actin was prepared from rabbit muscles according to Pardee and Spudich
\cite{par82} with an additional gel filtration step as suggested by
MacLean-Fletcher and Pollard \cite{mac80} using a Sephacryl S-300 HR column.
Tropomyosin/Troponin were prepared from the residue of rabbit muscle acetone
powder left after the actin extraction \cite{spu71}, and separated into
tropomyosin and troponin by hydroxyl apatite column chromatography
\cite{eis74}.  The purity of the proteins was checked by SDS polyacrylamide gel
electrophoresis \cite{lam70} stained with Commassie Blue, and estimated to be
of at least 95\% purity.  Actin was tested for its ability to polymerize by low
shear viscometry with the falling ball capillary apparatus as described by
Pollard and Cooper \cite{pol82}, and by fluorescence increase of 5\%
NBD-labeled actin \cite{det81}.  Functionality of tropomyosin/troponin in the
presence and absence of Ca$^{++}$ was characterized by an actin binding test:
they were added to F-actin in varying concentration ratios, centrifuged at 100
000 g for 1 h, and analyzed by SDS gel electrophoresis.

Actin was stored in a buffer containing 2 mM imidazole, 0.2 mM ATP, 0.2 mM DTT,
0.2 mM CaCl$_2$ and 0.05 vol.\% NaN$_3$. For the polymerization of actin, a
buffer with 2 mM imidazole, 0.5 mM ATP, 2 mM MgCl$_2$, 100 mM KCl, 0.2 mM DTT
and 0.2 mM CaCl$_2$ was used. The buffers where adjusted to a pH of 7.4. For
the experiment without Ca$^{++}$ the CaCl$_2$ was left out and 1 mM EGTA was
added. The molar ratio of actin:tropomyosin:troponin was 7:1:1.

For QELS measurements, all solutions were freed from dust by sterile
filtration, mixed, and stored overnight at $4^{\circ}$ C to achieve an
equilibrium polymerisation state. Samples to be inspected by electron
microscopy were adsorbed on glow discharged carbon coated formvar films on
copper grids, and negatively stained with uranyl acetate.

The experimental setup for QELS has been described in detail previously
\cite{sch89,pie92}. We used the correlator ALV 3000 (ALV Langen) with 1024
linear channels to calculate the dynamic structure factor. The light source was
an Innova 70-4 argon-ion laser from Coherent with 200 mW for the 488 nm line.

\section{Results}\label{exp}
We now turn to a discussion of the experimental data and their analysis in
terms of the theory described above.

\subsection*{Stretched Exponential}
\begin{figure}[tb]
  \narrowtext \hspace{0.005\columnwidth} \epsfxsize=0.9\columnwidth
  \epsfbox{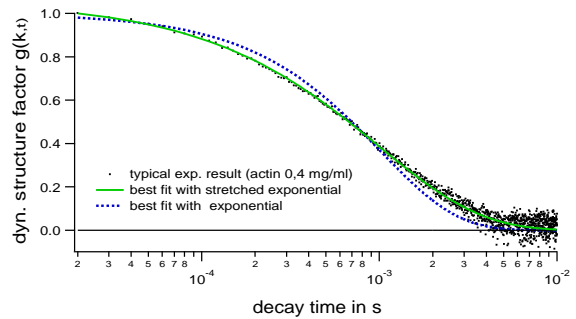} \vspace{0.2truecm}
\caption{Fit of theoretical dynamic structure factor Eq.\
  \protect\r{res} to experimental data for $k= 24.2 \, \mu {\rm m}^{-1}$.
  Clearly Eq.\ \protect\r{res} describes very well the experimental situation
  in the time interval $10^{-5} \sim 10^{-2}$ s, wheras the simple exponential
  fit is ruled out.}\label{fit}
\end{figure}

Figure~\ref{fit} shows a fit of the theoretical dynamic structure factor, Eq.\
\r{res}, to experimental data for a scattering angle of $90^{\circ}$
corresponding to $k= 24.2 \, \mu \rm m^{-1}$.  Note that there is only one free
parameter, $\gamma_k$.  An excellent fit of the experimental curves is obtained
in the time domain $10^{-5}\sim 10^{-2}$ s, for which the condition, Eq.
\r{tim}, is approximately fulfilled (cf.\ Footnote \protect\ref{taux}), whereas
a simple exponential decay is clearly ruled out. Hence the theory is well
suited to interpret our data within the present experimental accuracy.

\subsection*{Bending Modulus}
\begin{figure}[tb]
  \narrowtext \hspace{0.005\columnwidth} \epsfxsize=0.9\columnwidth
  \epsfbox{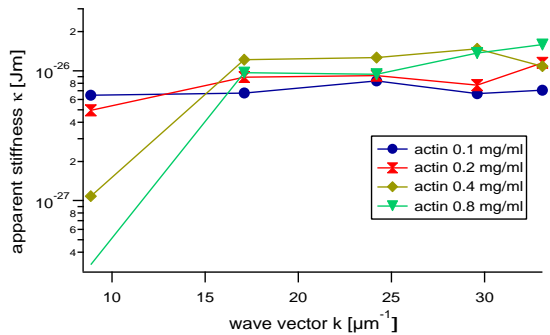} \vspace{0.2truecm}
\caption{Values for the
  bending modulus $\kappa$ obtained for various actin concentrations $c_a$ in
  the semidilute regime using Eq.\ \protect\r{res} and $\Lambda =\xi$.  Only
  large wave vectors $k\gg \xi^{-1}$ resolve single filaments and are
  accessible to our theory.}\label{con}
\end{figure}

As pointed out in Section~\ref{str} the decay rate $\gamma_k$, Eq.\ \r{dec}, is
determined by the bending modulus $\kappa$ and also by the screening length
$\Lambda$.  In order to check the effect of the screening length on the decay
rate, we measured the dynamic structure factor for various actin concentrations
$c_a$. The screening length was taken to be equal to the mesh size $\xi$, which
was found previously \cite{sch89} to obey the scaling law $\xi\, [\mu \rm m]
= 0.35\sqrt{c_a \, [\frac{\rm mg}{\rm ml}]}$.  The results are shown in
Figure~\ref{con}.  As discussed in Section~\ref{ins}, the theory only applies
to scattering vectors large enough to resolve single filaments.  Scattering
vectors $k$ smaller than the inverse mesh size $\xi^{-1}$ average over several
filaments.  Hence, the decay of the structure factor obeys Eq.\ \r{dec} only
for large scattering vectors $k$.  The onset of the deviation may be taken as a
lower bound for the mesh size.

Although the concentration was increased by a factor of 8 the derived values
for the bending modulus $\kappa$ are fairly consistent. They agree within a
standard deviation of 35\% for the entire range of scattering vectors and
concentrations. However, for a single scattering vector and fixed concentration
the deviations are considerably smaller, e.g.\ for $k=24.2 \, \mu \rm m^{-1}$
($90^{\circ}$) and $0.4 \, \frac{\rm mg}{\rm ml}$ the data are reproducible
within 5\%.  Considering the discussion at the end of Section \ref{str} and the
dependence of the value of the bending modulus $\kappa$ derived from Eq.\
\ref{res} on the hydrodynamic aspect ratio $\Lambda/a$ -- and hence on our
choice of the screening length $\Lambda$ -- we are presently not able to
determine an accurate absolute value for $\kappa$. (An average value of $9.5
\cdot 10^{-27}$ Jm was obtained for $\kappa$.) On the other hand, relative
changes in the stiffness can be resolved rather precisely.

Methods to determine $\Lambda$
experimentally are discussed in Section~\ref{dis}.

\subsection*{Effect of Tm/Tn Binding}

\begin{figure}[tb]
  \narrowtext \hspace{0.005\columnwidth} \epsfxsize=0.9\columnwidth
  \epsfbox{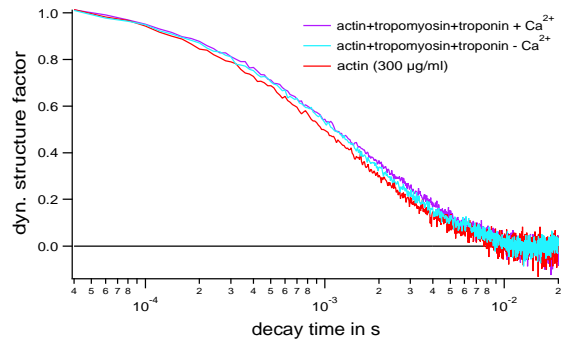}
\vspace{0.2truecm}
\caption{Effect of Tm/Tn binding with Ca$^{++}$ (0.2 mM uppermost curve) and
  without Ca$^{++}$ (closely below) in comparison with the pure actin sample
  (lower curve) observed at a scattering angle of $90^{\circ}$, temperature
  $T=10^{\circ}$ C and actin concentration $c_a=0.3 \, \frac{\rm mg}{\rm
    ml}$.}\label{sti}
\end{figure}

Figure~\ref{sti} shows the effect of the tropomyosin/troponin complex (Tm/Tn)
on the decay of the dynamic structure factor for actin solutions with a
concentration of $c_a= 0.3 \, \frac{\rm mg}{\rm ml}$ ($\simeq7.1\, \mu$M) at
$10^{\circ}$ C.  Since Ca$^{++}$ is known to regulate the coupling of the
Tm/Tn-complex to actin, experiments were performed with and without Ca$^{++}$.
Figure~\ref{sti} clearly shows the appreciable decrease of the decay rate in
the presence of Tm/Tn. The reduction of Ca$^{++}$ appears to decrease the
stiffness slightly but the effect is too weak to be considered significant with
the present accuracy of measurement.  The experiments were repeated several
times with different actin preparations and at two different temperatures
($10^{\circ}$ C and $25^{\circ}$ C) and the same absolute value of $\kappa$
as well as the same degree of stiffening by Tm/Tn was always observed.
The results of the measurements at $10^{\circ}$C are summarized in
Table~\ref{tro}.
$$\vbox{
\begin{table}[tb]
\begin{tabular}{|l|*{3}{r@{.}l|}}
\nobreak
scattering vector $k$ [$\mu\rm m^{-1}$] & {17}&1 & {24}&2 & {29}&6 \\
\hline
$\kappa$ for actin [$10^{-27}$ Jm] &  9&9 & 8&7 & 8&6 \\
\hline
$\kappa$ for actin + Tm/Tn $-{\rm Ca}^{++}$ &  {11}&4  & {12}&9 &  {14}&7 \\
\hline
$\kappa$ for actin + Tm/Tn $+{\rm Ca}^{++}$  & {12}&1  & {15}&9 & {13}&9  \\
\end{tabular}
\vspace{0.2truecm}
\caption{Summary of values obtained for the bending modulus $\kappa$ [Jm]
of actin in the presence of Tm/Tn with and without Ca$^{++}$ at different
scattering angles in comparison to pure actin.}\label{tro}
\end{table}}$$
To exclude preparation artifacts and in order to check whether Tm/Tn has some
effect on the mesh size, which would in turn affect the hydrodynamic screening
length $\Lambda$ and thus the derived value of $\kappa$, the actin network was
examined by negative-staining EM for all samples.  We could not observe an
effect of Tm/Tn on the network structure.  In summary, we find that in the
presence of Ca$^{++}$ Tm/Tn causes an increase of the bending modulus of
F-actin by about 50\%.

\section{General Discussion}\label{dis}
The present analysis of the dynamic structure factor of semidilute entangled
actin solutions shows that QELS is a reliable tool for the study of the
internal dynamics of semiflexible polymers.  However, as may be seen from Eq.\
\r{dec}, the interpretation of experimental data is not entirely
straightforward. The expression for the dynamic structure factor \r{res} is
ambiguous with respect to the microscopic source of an observed change in the
decay rate.  Changes may be caused by variations in the intrinsic stiffness of
the filament as well as by variations of the mesh size of the network,
which in turn affect the screening length $\Lambda$.  As experiments with
$\alpha-$actinin show \cite{gottp}, even a local change in the mesh size, as
caused by such crosslinking proteins, results in a corresponding change of the
decay rate over the whole range of scattering vectors.  This is exactly what
one would expect from Eq.\ \r{dec}, if the main contribution to the scattered
light is attributed to the crosslinked clusters (seen in EMs), which contain
most of the filaments.

The main uncertainty in the quantitative predictions of the QELS method
described above presently arises from the uncertainty in the absolute value of
the hydrodynamic screening length $\Lambda$. Setting it equal to the mesh size
seems to be reasonable, but perhaps one could do better. There are various
methods to determine $\Lambda$.  The QELS method itself as described above
could be used, if $\kappa$ was known. Direct information on $\Lambda$ should be
provided by the autocorrelation of a labeled filament in solution, which may be
investigated by fluorescence microscopy in the case of actin.  Another approach
exploits reptation. In the same manner as the transversal friction, Eq.\
\r{sim}, one may introduce the kinetic coefficient of the longitudinal center
of mass diffusion of semiflexible filaments in entagled networks
$\zeta_{\|}=2\pi \eta L/\log(\Lambda/a)$.

There are also effects from the center of mass and rotational motion, which we
have neglected so far.  Because of the polydispersity of the actin solution,
some filaments shorter than the mesh size ($L<\xi$) that are free to rotate and
diffuse are always present. Their hydrodynamic correlation is not disturbed,
nor are they hindered by entanglement. The center of mass and rotational
dynamics of those short filaments may to a good approximation be described by
the theory for rigid rod molecules \cite{doi86} (for a more complete treatment
including bending motions see \cite{son91}).  For the time being, the accuracy
of the experimental data does not allow for a quantitative analysis of this
small effect.

Finally we want to compare our findings with results obtained by the
fluorescence technique.  The two methods can be considered complementary: They
probe adjacent wavevector and time regimes. The fluorescence technique is more
direct. QELS, on the other hand, avoids problems associated with potential
perturbations from fluorescence markers (for the case of actin and phalloidin
see Ref.\ \cite{bre91}) and gives much better statistics.  For the time being,
a significant discrepancy remains concerning the effect of Tm/Tn. Tm/Tn was
seen to cause a softening of the filaments with the fluorescence technique
\cite{kas93}. This contradicts the QELS data given above as well as some recent
direct measurements \cite{koj94,isa95}.  We do not have a fully convincing
explanation for this discrepancy. A stiffening of actin by Tm/Tn binding seems
very probable, since other proteins binding along actin filaments are also
known to enhance the stiffness of the filaments, e.g.\ talin \cite{rud93}, a
protein involved in the membrane binding of actin in cells.

However, one may also think of a more intriguing scenario in terms of torsional
modes and other nonlinear or higher order derivative contributions to the
configurational energy of the filament.  No thorough theoretical treatment of
all those possible contributions has been given so far, and it is not known,
which of them are actually relevant. A first principles analysis of the various
linear modes of actin (in vacuo) has only been published very recently
\cite{avr95}. Especially for F-actin local torsional modes and additional
``groove swinging" and axial slipping motions are thought to be important
because of its double-stranded structure. Very strong evidence for local
torsional excitations was provided by high resolution electron microscopy
studies \cite{bre91} showing that the twist angle is not fixed but may
fluctuate by $\pm 10^{\circ}$. Torsional fluctuations of some 100 nm length
have been identified. The presence of torsional modes may affect the dynamic
structure factor in two ways: First, they cause ``crankshaft" motions of bent
chains and thus provide an additional mechanism contributing to the time decay
of the correlation function. Second, they could lead to a scale dependent
renormalized bending stiffness of actin as was recently observed for the
railway track model \cite{eve94}. Though we did not see any systematic
$k$dependence of the bending modulus with the QELS method, such effects may
well occur in other $k$regimes.  The more ``exotic" motions of actin are of
great biological interest, since they could play an active role in the
conformational changes of the actin-myosin-complex during ATP-cleavage. It is
hoped that future QELS studies with enhanced precision and larger ranges of
scattering vectors and measurement times will help to clarify this important
point.

\vspace{1cm}

Acknowledgment: This work was supported by the Deutsche
Forschungsgemeinschaft (Sa 246/24-1 and Fr 850-2) and in part by the National
Science Foundation under Grant No PHY 89-04035.  One of the authors (E.
Sackmann) gratefully acknowledges the hospitality encountered during his stay
at the Institute for Theoretical Physics, UCSB, under the directorship of J.
Langer. We are grateful to Toni Maggs, Kurt Kremer and Joseph K\"as for
enlightening discussions and for making available unpublished results. We thank
Irene Sprenger, who contributed to this work by making many EM-micrographs, and
our biochemical laboratory staff for the preparation of the proteins.

\end{multicols}

\end{document}